\documentstyle[pre,aps,epsfig]{revtex} 
\begin{document}

\draft

\title{Pattern and wavenumber selection in ferrofluids}

\author{Ren\'{e} Friedrichs and Andreas Engel}
\address{FNW/ITP, Otto-von-Guericke-Universit\"{a}t, Postfach 4120,
         D-39016 Magdeburg, Germany}
         
\date{\today}

\maketitle

\begin{abstract}
The formation of patterns of peaks on the free surface of a ferrofluid subject
to a magnetic field normal to the undisturbed interface is investigated
theoretically. The relative stability of ridge, square, and hexagon
planforms is studied using a perturbative energy minimization procedure.
Extending previous studies the finite depth of the fluid layer
is taken into account. Moreover, adding the wavenumber modulus $k$ to the set
of variational parameters also the wavenumber selection problem is
addressed. The results are compared with previous investigations and recent
experimental findings. 
\end{abstract}
 

\section{Introduction}

When a ferrofluid layer is subjected to a vertically oriented and uniform
magnetic field, above a critical value of the field strength a hexagonal
pattern of peaks appears on the surface of the liquid. This Rosensweig
or normal field instability was first observed by Cowley and Rosensweig
in 1967 \cite{Cowley,Rosensweig}. Further increase of the magnetic field up to
a second threshold gives rise to a transition from the hexagonal to a square
planform  \cite{Allais,Abou}.

The arrangement of peaks resulting from the Rosensweig instability into
patterns of different geometry is just one particular example from an
impressive variety of pattern formation in physical systems \cite{CrHo}. It is
well known that although the instability threshold itself can be obtained from
a linearized version of the underlying equations, the pattern selection problem
requires the inclusion of non-linear terms. A standard procedure to probe the
non-linear regime perturbatively is by means of amplitude equations
\cite{ampl,CrHo}. 

However, unlike many other examples of pattern forming physical systems
discussed in the literature the surface profile of a ferrofluid in a static
magnetic field is an equilibrium structure. Accordingly the relative stability
of planforms and possible transitions between different patterns can be
investigated theoretically by studying the appropriate thermodynamical
potential. Still the problem is a complicated non-linear task since the local
magnetic field determining the surface profile in turn depends on the surface
deflection via boundary conditions. As a consequence the variational
minimization of the thermodynamic potential in the surface profile cannot be
accomplished exactly and one has to resort to approximate procedures
applicable in the experimentally relevant situation of slightly overcritical
magnetic fields. 

In classical investigations along these lines Gailitis \cite{Gailitis} and 
Kuznetsov and Spektor \cite{Spektor} analyzed the stability of the different
patterns of the Rosensweig instability by means of an energy minimization
principle. These as well as related investigations using methods of functional
analysis \cite{Twombley,Silber} or a generalized Swift-Hohenberg equation
\cite{Herrero,Kubstrup} were confined to fluid layers of infinite depth.

On the other hand experimental investigations are usually done with rather
thin fluid layers and the effects of finite thickness on the {\it linear}
regime have been studied recently in great detail. For example the dispersion
relation of surface waves on a ferrofluid layer of arbitrary thickness was
determined in \cite{Wesfreid} and the influence of viscosity on the linear
dynamics was elucidated in \cite{Lange}. Taking advantage of the dependence of
the threshold of instability and the wavelength of the most unstable mode on
the thickness of the fluid layer in a clever way, it is e.g. possible to
measure both the normal and the anomalous dispersion branch of surface waves
on ferrofluids \cite{Mueller}. 

In the present paper we complement these investigations by a thorough study
of the weakly non-linear regime of the Rosensweig instability slightly above
the critical magnetic field for a ferrofluid of arbitrary depth. Taking the 
limit of infinite layer thickness we also critically discuss the
classical findings obtained in \cite{Gailitis,Spektor}. Moreover we are able to
quantify the restriction to sufficiently small susceptibilities $\chi$ of the
fluids which was always used in previous studies. 

Our method of investigation is a generalization of the variational
minimization of an energy functional already used in \cite{Gailitis,Spektor}. 
Near the instability this functional may be written as a power series in the
amplitude of the surface deflection and the minimization can be performed
explicitly. Moreover our approach allows the theoretical investigation of
the wavenumber selection problem addressed also in recent 
experiments \cite{Abou}. Including the wavenumber $k$ into the set of
variational parameters we determine the dependence of the wavenumber of the
patterns on the magnetic field and investigate the influence of a varying
wavenumber on the stability of hexagons and squares. Therefore by a slight
extension of our method we are able to study questions which are not easily
accessible to other non-linear methods, as e.g. amplitude equations. 

The paper is organized as follows. In section \ref{sec:formulation} the basic
equations are collected and transformed into a form suitable for the
calculation of the energy. In section \ref{sec:ansatz} a perturbation ansatz
for the surface deflection is put forward and the important issue of its
consistency and region of validity is discussed. Section \ref{sec:infinite}
is devoted to the comparison of our findings with the classical
results of pattern selection in a fluid with infinite depth. Subsequently in
section \ref{sec:finite} we consider a ferrofluid layer with arbitrary
thickness and analyze the effects of the finite depth on the stability of the
patterns. In section \ref{sec:wavenumber} we address the wavenumber selection
problem. Finally section \ref{sec:discussion} contains a summary and compares 
our findings with recent experimental results.

\section{Basic equations}
\label{sec:formulation}

\begin{figure}
\begin{center}
\mbox{\epsfxsize=90mm%
\epsffile[0 0 380 240]{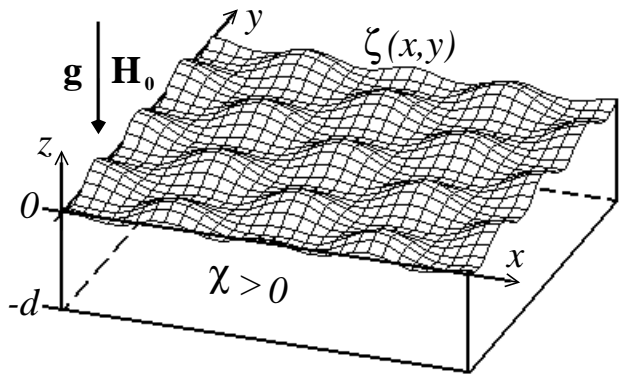}}
\end{center}
\caption{Schematic plot of a ferrofluid layer with infinite horizontal
  extension in an external magnetic field ${\bf H}_{0}$ parallel to the
  gravitational acceleration ${\bf g}$. We investigate the pattern formation  
 on the free surface $z=\zeta(x,y)$ for arbitrary depth $d$ of the magnetic
 fluid with susceptibility $\chi>0$.}
\label{fig:setup}
\end{figure}

We consider the situation sketched in Fig.\ \ref{fig:setup}.
A horizontally unbounded ferrofluid layer is subjected to an external magnetic 
field ${\bf H}_{0}$ which in the absence of any magnetic permeable material is
of the form ${\bf H}_{0}(x,y,z)=-H_{0}{\bf e}_{z}$. The incompressible
magnetic fluid of density $\rho$, surface tension $\sigma$ and susceptibility
$\chi$ is bounded from below at $z=-d$ by an impermeable material and has a
free surface described by $z=\zeta(x,y)$ with the magnetically impermeable air
above. The gravitational acceleration ${\bf g}=-g{\bf e}_{z}$ acts parallel to
the z-axis.  
Our aim is to determine which static profile $\zeta(x,y)$ develops for a 
magnetic field ${\bf H}_{0}$ strong enough to destabilize the flat surface
$\zeta(x,y)=0$. 

Stable configurations of the ferrofluid surface with infinite horizontal
extension are given by minima of the thermodynamic potential per unit area in 
the $x-y$ plane
\begin{equation}
f[\zeta(x,y)]=\left\langle 
\frac{\rho g}{2} \, \zeta^{2}(x,y) -
\frac{\mu_{0}}{2} \!\!\! \int\limits_{-d}^{\zeta(x,y)} \!\!\!\! dz \,
{\bf H}_{0} \chi {\bf H}(x,y,z) +
\sigma \sqrt{1+(\partial_{x}\zeta(x,y))^2+(\partial_{y}\zeta(x,y))^2} 
\right\rangle.
\label{eq:tildef}
\end{equation}
Here $\mu_{0}$ is the permeability of free space, ${\bf H}(x,y,z)$ is the
magnetic field in the presence of the ferrofluid, and the brackets denote the
average over the $x-y$ plane defined by  
\begin{equation}
\bigl\langle F(x,y) \bigr\rangle := \lim_{L \to \infty}
\frac{1}{4 L^{2}} \!\int\limits_{-L}^{L} \!\! dx \!\int\limits_{-L}^{L} \!\!
dy \, F(x,y).
\label{eq:average}
\end{equation}
The three terms on the r.h.s.\ of Eq.\ (\ref{eq:tildef}) describe the 
hydrostatic energy, the magnetic energy \cite{Landau}, and the surface
energy respectively. Note that both the hydrostatic and the surface energy
increase when the interface profile starts to deviate from the flat reference
state, whereas the magnetic energy decreases. For sufficiently large 
${\bf H}_{0}$ this gives rise to the normal field or Rosensweig instability
\cite{Cowley,Rosensweig}.

The magnetic fields ${\bf B}$ and ${\bf H}$ are determined by the static
Maxwell equations,  
\begin{equation}
\nabla \cdot {\bf B} = 0 \quad \mathrm{and} \quad
\nabla \times {\bf H} = {\bf 0},
\label{eq:Maxwell}
\end{equation}
together with the boundary conditions,
\begin{equation}
\lim_{z \to \pm\infty} {\bf H}(x,y,z) = - H_{0}{\bf e}_{z},
\label{eq:condition}
\end{equation}
\begin{equation}
\left[({\bf B}_{\mathrm{a}}-{\bf B}_{\mathrm{ff}})\cdot {\bf n}\right]
{\Big |}_{z=\zeta}= 0 \quad \mathrm{and} \quad
\left[({\bf H}_{\mathrm{a}}-{\bf H}_{\mathrm{ff}})\times{\bf n}\right]
{\Big |}_{z=\zeta}= {\bf 0},
\label{eq:abound}
\end{equation}
\begin{equation}
\left[({\bf B}_{\mathrm{ff}}-{\bf B}_{\mathrm{b}})\cdot {\bf e}_{z}\right]
{\Big |}_{z=-d}= 0 \quad \mathrm{and} \quad
\left[({\bf H}_{\mathrm{ff}}-{\bf H}_{\mathrm{b}})\times{\bf e}_{z}\right]
{\Big |}_{z=-d}= {\bf 0},
\label{eq:bbound}
\end{equation}
where ${\bf n}$ denotes the normal vector on the surface $\zeta(x,y)$ and 
the respective subscripts indicate the fields above, in and below the
ferrofluid. The minimization problem for the thermodynamic potential becomes
non-trivial since the magnetic field ${\bf H}(x,y,z)$ depends on the profile 
$\zeta(x,y)$ of the surface.

Throughout the paper we will assume the linear relation 
\begin{equation}
{\bf B} = \mu_{0}(1+\chi){\bf H}
\label{eq:relation}
\end{equation}
between the magnetic induction and the magnetic field. For the experimentally
relevant magnetic fluids and fields \cite{Abou} this approximation deviates
about 5\% from the exact relation between ${\bf B}$ and ${\bf H}$.

It is convenient to introduce a scalar magnetic potential $\psi(x,y,z)$
defined by 
\begin{equation}
{\bf H}= - \nabla \psi
\label{eq:psi}
\end{equation}
which by (\ref{eq:Maxwell}) has to satisfy the Laplace equation
\begin{equation}
\Delta \psi = 0.
\label{eq:Laplace}
\end{equation}

The two characteristic scales of the problem are the critical wavenumber 
at the onset of the instability,
\begin{equation}
k_{\mathrm{c},\infty} = \sqrt{\frac{\rho g}{\sigma}},
\label{eq:kcinf}
\end{equation}
for a ferrofluid with infinite depth $d \to \infty$ and the corresponding
critical magnetic field, 
\begin{equation}
H_{\mathrm{c},\infty} = \sqrt{\frac{(1+\chi)(2+\chi)2\sqrt{\rho g\sigma}}
{\chi^2 \mu_{0}}},
\label{eq:Hcinf}
\end{equation}
which were first derived in \cite{Cowley} from a linear stability analysis.
Henceforth we therefore measure all lengths in units of
the capillary length $k_{\mathrm{c},\infty}^{-1}$, all wavenumbers
in units of the critical wavenumber $k_{\mathrm{c},\infty}$, the
magnetic field ${\bf H}_{0}$ in units of the critical value
$H_{\mathrm{c},\infty}$, the scalar magnetic potential $\psi$ in units
of $H_{\mathrm{c},\infty}/k_{\mathrm{c},\infty}$, and energies per unit area in
units of $\sigma$. Moreover it is convenient
to introduce the following rescaled magnetic potentials in the space above, in
and below the ferrofluid respectively:
\begin{eqnarray}
\psi_{\mathrm{a}} &:=& \psi\;\frac{(2+\chi)}{H_{0}\chi}, \label{eq:psia} \\
\psi_{\mathrm{ff}}&:=& \psi\;\frac{(1+\chi)(2+\chi)}{H_{0}\chi},
\label{eq:psiff}\\ 
\psi_{\mathrm{b}} &:=& \psi\;\frac{(2+\chi)}{H_{0}\chi}. \label{eq:psib} 
\end{eqnarray}
Each of these potentials has to fulfill the Laplace equation
(\ref{eq:Laplace}). Using the abbreviation 
\begin{equation}
\eta:=\frac{\chi}{2+\chi},
\end{equation}
the asymptotic form of the magnetic field for $|z|\to \infty$ as specified by 
Eq.\ (\ref{eq:condition}) gives rise to the requirements 
\begin{equation}
\lim_{z \to +\infty} \partial_{z}\psi_{\mathrm{a}}(x,y,z) = \frac{1}{\eta} =
\lim_{z \to -\infty} \partial_{z}\psi_{\mathrm{b}}(x,y,z)
\label{eq:limits}
\end{equation}
for the magnetic potentials. Moreover the boundary conditions (\ref{eq:abound})
and (\ref{eq:bbound}) for the magnetic fields translate into the following
boundary conditions for the potentials:
\begin{equation}
\left[\partial_{x}\psi_{\mathrm{a}}-\partial_{x}\psi_{\mathrm{ff}}\right]
{\bigg |}_{z=\zeta}\partial_{x}\zeta +
\left[\partial_{y}\psi_{\mathrm{a}}-\partial_{y}\psi_{\mathrm{ff}}\right]
{\bigg |}_{z=\zeta}\partial_{y}\zeta -
\left[\partial_{z}\psi_{\mathrm{a}}-\partial_{z}\psi_{\mathrm{ff}}\right]
{\bigg |}_{z=\zeta}
=0,
\label{eq:ntop}
\end{equation}
\begin{equation}
\frac{1+\eta}{1-\eta}\psi_{\mathrm{a}}{\bigg |}_{z=\zeta}
-\psi_{\mathrm{ff}}{\bigg |}_{z=\zeta}=0,
\label{eq:ttop}
\end{equation}
\begin{equation}
\left[\partial_{z}\psi_{\mathrm{ff}}-\partial_{z}\psi_{\mathrm{b}}\right]
{\bigg |}_{z=-d}
=0,
\label{eq:nbot}
\end{equation}
\begin{equation}
\psi_{\mathrm{ff}}{\bigg |}_{z=-d}-
\frac{1+\eta}{1-\eta}\psi_{\mathrm{b}}{\bigg |}_{z=-d}=0.
\label{eq:tbot}
\end{equation}

Using Eqs.\ (\ref{eq:psi}) and (\ref{eq:psiff}) and exploiting the fact that 
${\bf H}_{0}$ is parallel to the z-axis, we finally get the energy $f$ as
functional of the surface deflection $\zeta =
\zeta(x,y)$ in the form
\begin{equation}
f[\zeta]=\left\langle
\frac{\zeta^2}{2} - H_{0}^2
\left(\psi_{\mathrm{ff}}{\big |}_{z=\zeta}
-\psi_{\mathrm{ff}}{\big |}_{z=-d}\right) + 
\sqrt{1+(\partial_{x}\zeta)^2+(\partial_{y}\zeta)^2} 
\right\rangle.
\label{eq:f}
\end{equation}
As stated above, the main problem, rendering a straightforward minimization of 
$f$ in $\zeta(x,y)$ impossible, is the rather implicit dependence of the
potential $\psi_{\mathrm{ff}}(x,y,z)$ on the surface deflection $\zeta(x,y)$ 
specified by the boundary conditions (\ref{eq:ntop}) and (\ref{eq:ttop}).

\section{The perturbation ansatz}
\label{sec:ansatz}

The variational problem for the energy functional posed in the last paragraph
can in general not be solved exactly. In order to make analytic progress, we
restrict ourselves to the vicinity of the critical magnetic field and assume
that the amplitude of the surface deflection is still small. It is then
possible to expand the energy in this amplitude and to retain only the first
terms. In this paper we will consider an expansion of the energy up to fourth
order in the amplitude. The applicability of this approach will be critically
discussed below. 

\begin{figure}
\begin{center}
\mbox{\epsfxsize=100mm%
\epsffile[48 257 511 517]{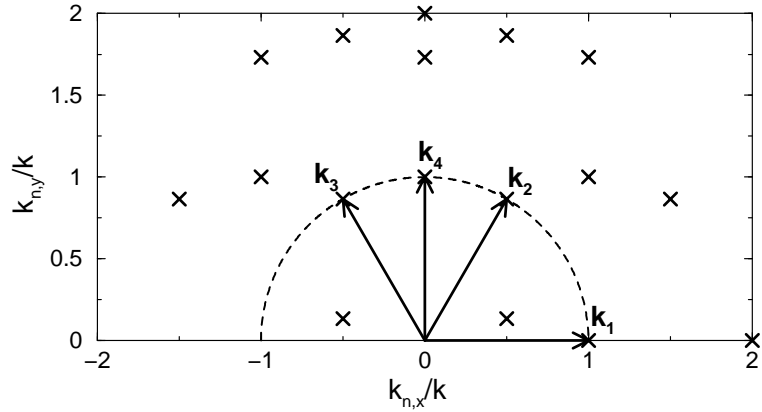}}
\end{center}
\caption{Wave vectors ${\bf k}_{1}$ to ${\bf k}_{4}$ of the four main modes in
  the perturbation ansatz (\ref{eq:pertan}) (arrows).  ${\bf k}_{4}$
is perpendicular to ${\bf k}_{1}$ and the angle between ${\bf k}_{1}$ and 
${\bf k}_{2}$ or ${\bf k}_{2}$ and ${\bf k}_{3}$ is $\pi/3$. The endpoints of
the wave vectors of the higher harmonics are represented by crosses. Each
of these wavevectors is the sum or difference of two of the main wavevectors.}
\label{fig:wavevectors}
\end{figure}

Generalizing the perturbation expansions used in
\cite{Gailitis,Spektor,Zaitsev,Engel}  
for the one-dimensional variant of the Rosensweig instability, we write the
surface profile in the form
\begin{equation}\label{eq:pertan}
\zeta(x,y)=\sum_{n=1}^{4}A_n\cos({\bf k}_n\cdot{\bf r})
+\sum_{n=5}^{17}B_n\cos({\bf k}_n\cdot{\bf r}),
\end{equation}
where ${\bf r}=(x,y)$ and ${\bf k}=(k_{x},k_{y})$ are two-dimensional vectors.
The terms with $n=1,...,4$ constitute the main modes. Their corresponding
wavevectors ${\bf k}_1$ to ${\bf k}_4$ have all the same modulus $k$ whereas
their mutual orientation is chosen such that the ansatz allows the description
of ridges (e.g. $A_{1}=A,\; A_{2}=A_{3}=A_{4}=0$), squares
(e.g. $A_{1}=A_{4}=A,\; A_{2}=A_{3}=0$), and hexagons
(e.g. $A_{1}=A_{2}=A_{3}=A,\; A_{4}=0$) (see Fig.\ \ref{fig:wavevectors}). 

The terms with $n=5,...,17$ are higher harmonics with wave vectors ${\bf k}_5$
to ${\bf k}_{17}$ being linear combinations of two 
wave vectors of the main modes and amplitudes $B_n$ of order $O(A^2)$. These
terms are needed to satisfy the minimum conditions for the energy functional
to the required order $O(A^3)$. The intuitive meaning of this fact is that
reliable results on the relative stability of different planforms requires
some information on the deviation of the nonlinear surface profile 
from the simple cosine shape describing the linear instability. On the other
hand there is no need to include even higher harmonics in the ansatz
(\ref{eq:pertan}) since the corresponding contributions would average to zero
in the $x-y-$integrations in the definition (\ref{eq:f}) of the
energy functional. In conclusion the chosen ansatz is the only consistent
perturbation ansatz for the energy which includes terms up to fourth order in
the amplitude of the surface deflection. The values of the amplitudes
$A_1,\dots, A_4$ and $B_5,\dots, B_{17}$ as well as the wave vector modulus
$k$ of the main modes are the free parameters which may be used to minimize
the energy. In particular the possible minimization in $k$ is a special
advantage of our approach since it allows an analysis of the non-linear wave
vector selection problem not easily accessible to other approaches.  

Because of the boundary conditions at the surface of the fluid
the magnetic potential $\psi$ must have a similar dependence on $x$ and $y$
as $\zeta(x,y)$. The ansatzes 
\begin{eqnarray}
\psi_{\mathrm{a}} & = &
\frac{z}{\eta}+
\sum_{n=0}^{17}u_n
\mathrm{e}^{-|{\bf k}_n|z}\cos({\bf k}_n\cdot{\bf r}),
\nonumber\\
\psi_{\mathrm{ff}} & = &
\frac{z}{\eta}+
\sum_{n=1}^{17}[v_n^{+}
{\mathrm e}^{|{\bf k}_n|z}+ v_n^{-}
\mathrm{e}^{-|{\bf k}_n|z}]\cos({\bf k}_n\cdot{\bf r}),
\nonumber\\
\psi_{\mathrm{b}} & = &
\frac{z}{\eta}+\frac{2d}{1+\eta}+
\sum_{n=0}^{17}w_n
\mathrm{e}^{|{\bf k}_n|z}\cos({\bf k}_n\cdot{\bf r}),
\label{eq:ansatz}
\end{eqnarray}
fulfill the Laplace equation\ (\ref{eq:Laplace}) and the asymptotic boundary
conditions (\ref{eq:limits}). 
Omitting the mode ${\bf k}_{0}:=(0,0)$ in the potential $\psi_{\mathrm{ff}}$
fixes the free constant of the potential $\psi$. Plugging
Eq.\ (\ref{eq:ansatz}) into the four remaining boundary conditions 
Eqs.\ (\ref{eq:ntop})-(\ref{eq:tbot}) we
can determine the parameters $u_n,v_n^{+},v_n^{-}$ and $w_n$ as functions of 
the surface deflection amplitudes by expanding the equations up to third order
in the $A_n$. The resulting expressions specify the dependence of the magnetic
field on the surface profile which is then used to calculate the energy $f$
using Eq.\ (\ref{eq:f}) up to fourth order in the amplitudes $A_n$. The
resulting dependence of $f$ on the higher order amplitudes $B_n$ is
simple and the minimization in the $B_n$ can be performed
explicitly. Subtracting from the resulting expression the reference value of
the energy for a flat interface we finally arrive at
\begin{eqnarray}\label{eq:defftilde}
f_{d}(\{A_{n}\},k) &:= & f(\{A_{n}\},k)-f(\zeta=0)\\
& = & -\frac{1}{2}l(\epsilon,k)[A_{1}^{2}+A_{2}^{2}+A_{3}^{2}+A_{4}^{2}]
-\gamma(\epsilon,k)[A_{1}A_{2}A_{3}]
+\frac{1}{4}g(\epsilon,k)[A_{1}^{4}+A_{2}^{4}+A_{3}^{4}+A_{4}^{4}]+
\nonumber\\ & & 
+\frac{1}{2}g_{h}(\epsilon,k)
[A_{1}^{2}A_{2}^{2}+A_{2}^{2}A_{3}^{2}+A_{3}^{2}A_{1}^{2}] 
+\frac{1}{2}g_{t}(\epsilon,k)[A_{2}^{2}A_{4}^{2}+A_{3}^{2}A_{4}^{2}]
+\frac{1}{2}g_{n}(\epsilon,k)[A_{1}^{2}A_{4}^{2}]
+O(A^{5}).
\end{eqnarray}
This expression gives the energy of a ferrofluid layer of depth $d$ with
surface profile $\zeta(x,y)$ as specified by (\ref{eq:pertan}) at arbitrary
strength of the external field $H_0$ up to fourth order in the deflection
amplitudes $A_n$. The dependence on $H_0$ is expressed via the
supercriticality parameter
\begin{equation}
\epsilon=\frac{H_{0}^{2}}{H_{\mathrm{c},d}^{2}}-1.
\end{equation}
The function $l(\epsilon,k)$ is given by 
\begin{equation}
l(\epsilon,k)=-\frac{1}{2}+
(1+\epsilon)H_{\mathrm{c},d}^{2}\frac{1-e^{-2kd}\eta}{1-e^{-2kd}\eta^{2}}k
-\frac{1}{2}k^{2}. 
\label{eq:l(epsilon,k)}
\end{equation}
It fixes the critical wavenumber $k_{\mathrm{c},d}$ for the onset of the
instability and the corresponding critical magnetic field $H_{\mathrm{c},d}$
for a ferrofluid with {\it finite} depth $d$ via the conditions 
\begin{mathletters}
\label{eq:Hcdkcd}
\begin{equation}
l(\epsilon=0){\big |}_{k=k_{\mathrm{c},d}}=0
\end{equation}
and 
\begin{equation}
\frac{\partial l(\epsilon=0)}{\partial k}{\big |}_{k=k_{\mathrm{c},d}}=0.
\end{equation}
\end{mathletters}

Determining the minima of $f$ with respect to the amplitudes $A_n$, one
finds that $\epsilon\to 0$ with $A_n\to 0$ as expected:
the external field will be near to its critical value if the
{\it equilibrium} values of the amplitudes $A_n$ are small.
For ridges and squares the explicit scaling is 
$\epsilon\sim A^2$ for $A\to 0$, in the case of hexagons one finds 
$\epsilon\sim -A$ for small $A$. This in turn implies that to the desired order
$O(A^4)$ of the expansion of $f$ in the amplitudes $A_n$ only the 
$\epsilon$-dependence of $l$ and $\gamma$ has to be retained. The exact form
of these dependencies is 
$l(\epsilon,k)=l(0,k)+\epsilon\partial l(\epsilon,k)/\partial \epsilon$
and $\gamma(\epsilon,k)=(1+\epsilon)\gamma(k)$. In this way we arrive at our
main result for the energy $f_d(\{A_{n}\},k)$ of the surface deflection which
describes the equilibrium configurations consistently up to fourth order in
the amplitudes $A_n$:
\begin{eqnarray}
f_d(\{A_{n}\},k) & = & -\frac{1}{2}l(\epsilon,k) 
[A_{1}^{2}+A_{2}^{2}+A_{3}^{2}+A_{4}^{2}]
-(1+\epsilon)\gamma(k)[A_{1}A_{2}A_{3}]
+\frac{1}{4}g(k)[A_{1}^{4}+A_{2}^{4}+A_{3}^{4}+A_{4}^{4}]+ \nonumber\\ & &
+\frac{1}{2}g_{h}(k)[A_{1}^{2}A_{2}^{2}+A_{2}^{2}A_{3}^{2}+A_{3}^{2}A_{1}^{2}]
+\frac{1}{2}g_{t}(k)[A_{2}^{2}A_{4}^{2}+A_{3}^{2}A_{4}^{2}]
+\frac{1}{2}g_{n}(k)[A_{1}^{2}A_{4}^{2}]
+O(A^{5}).
\label{eq:u}
\end{eqnarray}
The explicit expressions for the various coefficients in (\ref{eq:u}) are
rather long. We therefore display their form here only for the comparatively
simple situation of a fluid of infinite depth, $d\to \infty$:
\begin{mathletters}
\label{eq:coeffs}
\begin{equation}
\gamma_{\infty}(k)=\frac{3}{4}\eta k^{2},
\end{equation}
\begin{equation}
g_{\infty}(k)=\frac{1}{16}k^{3}(8-3k)-\frac{k^{4}\eta ^{2}}{1-4k+4k^{2}},
\end{equation}
\begin{equation}
g_{h,\infty}(k)=\frac{1}{4}k^{3}(
11\eta^{2}-7\eta^{2}\sqrt{3}+3\sqrt{3}-\frac{3}{4}k-3)-\frac{3}{8}\frac{k^{4}
\eta^{2}(19-8\sqrt{3})}{1-2\sqrt{3}k+3k^{2}}, 
\end{equation}
\begin{eqnarray}
g_{t,\infty}(k) & = &
k^{3}(\frac{3}{4}\sqrt{3}\sqrt{2}-1+4\eta^{2}-\frac{7}{4}\eta^{2} 
\sqrt{3}\sqrt{2}-\frac{5}{16}k)+ \nonumber\\ & & 
-\frac{1}{8}\frac{\eta^{2}k^{4}(18+\sqrt{3}-4\sqrt{2}-4\sqrt{3}\sqrt{2})}
{2-k\sqrt{3}\sqrt{2}+k^{2}+k\sqrt{2}-\sqrt{3}}
-\frac{1}{8}\frac{\eta^{2}k^{4}(18-\sqrt{3}+4\sqrt{2}-4\sqrt{3}\sqrt{2})}
{2-k\sqrt{3}\sqrt{2}+k^{2}-k\sqrt{2}+\sqrt{3}},
\end{eqnarray}
\begin{equation}
g_{n,\infty}(k)=k^{3}(-\frac{1}{8}k-3\eta^{2}\sqrt{2}-1+4\eta^{2}
+\sqrt{2})-\frac{k^{4}\eta^{2}(9-4\sqrt{2})}{1-2k\sqrt{2}+2k^{2}}.
\end{equation}
\end{mathletters}

\section{Ferrofluid layer with infinite depth}
\label{sec:infinite}

In this section we consider a ferrofluid of infinite depth $d \to \infty$
and fix the wavenumber modulus $k$ at its critical value
$k_{\mathrm{c},\infty}=1$. This enables us to compare our findings with the
well-known results of previous theoretical investigations.

Possible surface patterns together with their amplitudes are given by
stationary points of the energy function $f_{\infty}(\{A_n\},k=1)$. We will
in particular study ridges (``rolls'') given by $A_{1}=A_{R}$ with 
$A_{2}=A_{3}=A_{4}=0$, squares represented by $A_{1}=A_{4}=A_{S}$ with 
$A_{2}=A_{3}=0$, and hexagons described by $A_{1}=A_{2}=A_{3}=A_{H}$ with 
$A_{4}=0$. The last planform is shown in Fig.\ \ref{fig:setup}. Using 
Eqs.\ (\ref{eq:l(epsilon,k)}), (\ref{eq:u}) and (\ref{eq:coeffs})
we find for the corresponding amplitudes 
\begin{equation}
A_{R}(\epsilon)=\sqrt{\frac{\epsilon}{g_{\infty}}},
\label{eq:AR}
\end{equation}
\begin{equation}
A_{S}(\epsilon)=\sqrt{\frac{\epsilon}{g_{n,\infty}+g_{\infty}}},
\label{eq:AS}
\end{equation}
and 
\begin{equation}
A_{H}(\epsilon)=\frac{1}{2}
\frac{\gamma_{\infty}(1+\epsilon)+\sqrt{\gamma_{\infty}^{2}
    (1+\epsilon)^{2}+4\epsilon(2g_{h,\infty}+g_{\infty})}}
    {2g_{h,\infty}+g_{\infty}}
\label{eq:AH}
\end{equation}
respectively. Together with $\chi$ also $\gamma$ is always
positive\footnote{This holds true also in the case of a finite layer depth
  $d$.}. Hence only up-hexagons with a peak instead of a dip in the center are
stable. 

Our results for the hexagons differ slightly from the classical ones reported
in \cite{Gailitis} in which the dependence of the cubic term in Eq.\
(\ref{eq:u}) on $\epsilon$ is missing. In \cite{Gailitis,Spektor} this 
dependence was omitted because the discussion was restricted
to fluids with $\gamma \ll 1$ so that the approximate scaling $\epsilon\sim
A^{2}$ is very similar to the true one $\epsilon\sim -\gamma A+ O(A^{2})$.
Otherwise our results coincide with those of Gailitis, in particular using
$k=k_{\mathrm{c},\infty}=1$ in (\ref{eq:coeffs}) the respective
expressions found in \cite{Gailitis} are reproduced. The investigations of
Kuznetzov and Spektor \cite{Spektor} were restricted to first order in $\eta$
and are contained in both our results and those of Gailitis. 

\begin{figure}
\begin{center}
\mbox{\epsfxsize=100mm%
\epsffile[20 37 507 460]{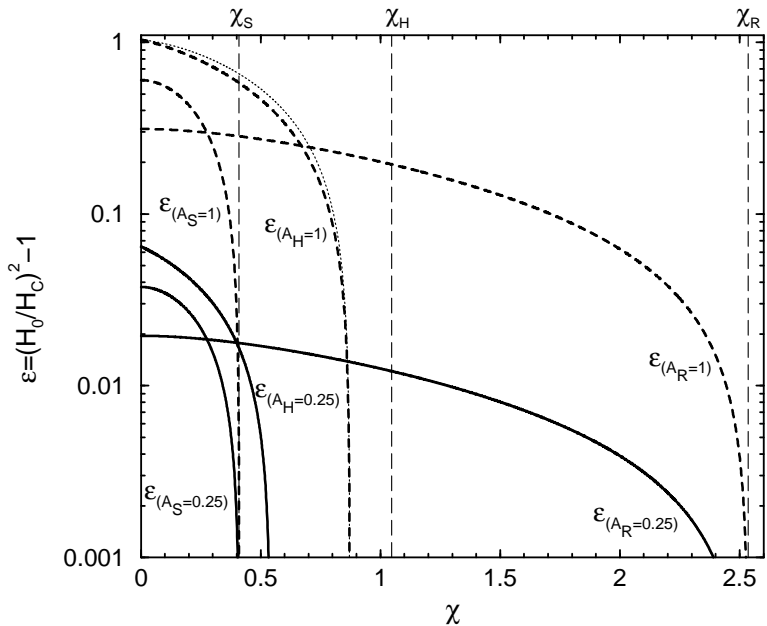}}
\end{center}
\caption{Lines separating regions in the $\chi-\epsilon$ plane in which the
  amplitudes of  ridges $A_{R}$, squares $A_{S}$ and hexagons $A_{H}$ are
  smaller than $0.25$ (full lines) or smaller than $1$ (dashed lines). 
  The thin dotted line represents the results given in \protect\cite{Gailitis}.
  For susceptibilities larger than the critical values $\chi_{R},\chi_{S}$ and
  $\chi_{H}$ the amplitudes diverge in the considered approximation 
  (\ref{eq:pertan}).} 
\label{fig:validation}
\end{figure}

Before addressing the relative stability of the different planforms introduced
above it is important to characterize the domain of validity of our
central expression (\ref{eq:u}) for the energy. Being a perturbative result
the corresponding amplitudes $A_n$ must not be too
large. Fig.\ \ref{fig:validation} gives a quantitative impression of what that
means by displaying lines in the $\chi-\epsilon$ plane corresponding to fixed
values of the amplitudes of different patterns. It is in particular important
to realize that for sufficiently large values of the susceptibility $\chi$ the
fourth order coefficients in the energy functional may change sign. The
corresponding patterns than appear through a backward bifurcation
\cite{Silber}. At the same time {\it higher order} terms (i.e. sixth or higher
order in the amplitudes $A_n$) in the energy are
necessary to saturate the instability. The concrete values of $\chi$ for the
planforms considered are $\chi_{R}\approx 2.54$ ($g_{\infty}$ becomes
negative), $\chi_{S}\approx 0.41$ ($g_{n,\infty}+g_{\infty}$ becomes
negative), and  $\chi_{H}\approx 1.05$ ($2g_{h,\infty}+g_{\infty}$ becomes
negative). Beyond these values of $\chi$ our ansatz (\ref{eq:pertan}) is 
unable to describe the arising pattern. The value of $\chi_{R}$ was first 
calculated in \cite{Zaitsev}.

Keeping this limitation of our approach in mind we have investigated the 
pattern selection problem by studying the character of the extremum of the
energy functional (\ref{eq:u}) at the ridge, square and hexagon solutions
given in Eqs.\ (\ref{eq:AR})-(\ref{eq:AH}). This is rather similar to a linear
stability analysis of the fixpoint solutions of the corresponding amplitude
equations \cite{Cil}. The results are as follows:
Ridges are never stable, since $g_{n,\infty}<g_{\infty}$ always.
We have always $g_{\infty}+g_{n,\infty}<g_{h,\infty}+g_{t,\infty}$
and $g_{n,\infty}<g_{\infty}$. Hence (for $\chi<\chi_{S}$) squares are stable
if $\displaystyle\epsilon>\epsilon_{S}=
\frac{(g_{\infty}+g_{n,\infty}-g_{h,\infty}-g_{t,\infty}
+\sqrt{(g_{\infty}+g_{n,\infty}-g_{h,\infty}-g_{t,\infty})^{2}
-4\gamma_{\infty}^{2}(g_{\infty}+g_{n,\infty})})^{2}} 
{4\gamma_{\infty}^{2}(g_{\infty}+g_{n,\infty})}$.  
In order for hexagons to be stable (for $\chi<\chi_{H}$) the following
three conditions have to be fulfilled.
\begin{itemize}
\item 
$\displaystyle\epsilon>\epsilon_{F}= 
\frac{(\gamma_{\infty}^{2}\!\!+2g_{h,\infty}\!\!+g_{\infty}\!
-\!\sqrt{(\gamma_{\infty}^{2}\!\!+2g_{h,\infty}\!\!+g_{\infty})
(2g_{h,\infty}\!\!+g_{\infty})})
(2g_{h,\infty}\!\!+g_{\infty}\!
-\!\sqrt{(\gamma_{\infty}^{2}\!\!+2g_{h,\infty}\!\!+g_{\infty})
(2g_{h,\infty}\!\!+g_{\infty})})}
{\gamma_{\infty}^{2}\sqrt{(\gamma_{\infty}^{2}+2g_{h,\infty}+g_{\infty})
(2g_{h,\infty}+g_{\infty})}}$,
\item either  
$\displaystyle\epsilon<\epsilon_{H}=
\frac{\gamma_{\infty}^{2}(g_{h,\infty}+2g_{\infty})}
{(\gamma_{\infty}^{2}+g_{h,\infty}-g_{\infty})(g_{h,\infty}-g_{\infty})}$
or $g_{h,\infty}<g_{\infty}$,  
\item either
$\displaystyle\epsilon<\epsilon_{h}= 
\frac{\gamma_{\infty}^{2}(g_{n,\infty}+2g_{t,\infty})} 
{(\gamma_{\infty}^{2}\!
+2g_{h,\infty}\!+g_{\infty}\!-2g_{t,\infty}\!-g_{n,\infty})
(2g_{h,\infty}\!+g_{\infty}\!-2g_{t,\infty}\!-g_{n,\infty})}$  
or $2g_{h,\infty}+g_{\infty}<2g_{t,\infty}+g_{n,\infty}$. 
\end{itemize}
 
\begin{figure}
\begin{center}
\mbox{\epsfxsize=157mm%
\epsffile[2 16 594 337]{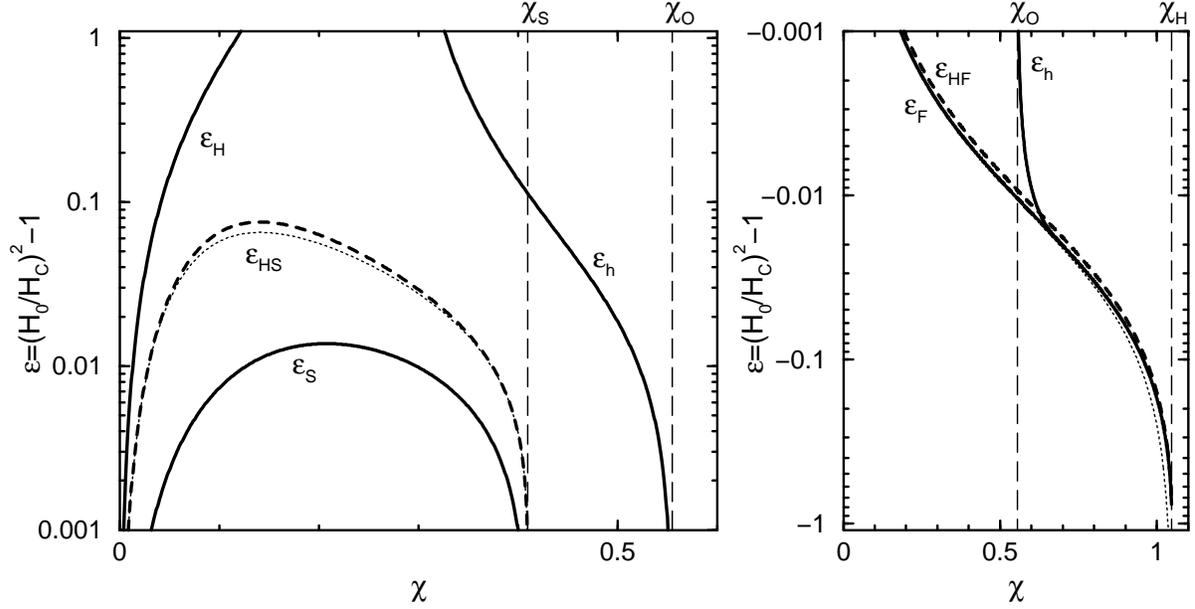}}
\end{center}
\caption{Limits of the stability regions (solid lines) and Maxwell points 
(dashed lines) in the $\chi-\epsilon$ plane. The left figure refers to positive
values of $\epsilon$, the right to the subcritical region $\epsilon<0$.  
Squares are stable if $\epsilon>\epsilon_{S}$. Hexagons are stable
if $\epsilon<\epsilon_{H}$, $\epsilon<\epsilon_{h}$
and $\epsilon>\epsilon_{F}$. The two thin dotted lines illustrate
exemplarily the corresponding findings if the dependence of the cubic term in
Eq.\ (\ref{eq:u}) on $\epsilon$ is neglected.} 
\label{fig:stability}
\end{figure}

These results are displayed in Fig.\ \ref{fig:stability}. For regions in which
two different patterns are stable, i.e. in which different local minima of
the energy exist, we have also determined the respective Maxwell values of
$\epsilon$ at which these minima have the same value. For
$\epsilon=\epsilon_{HS}$ the energy of the hexagon pattern is 
equal to the energy of the square pattern, for
$\epsilon=\epsilon_{HF}$ it is equal to the energy of the flat
surface.

For all susceptibilities either the relation $g_{h,\infty}<g_{\infty}$
or the relation $2g_{h,\infty}+g_{\infty}<2g_{t,\infty}+g_{n,\infty}$ is
violated and consequently hexagons always become unstable for sufficiently
large $\epsilon$. 

Of particular importance is the line denoted by $\epsilon_h$. For
$\epsilon>\epsilon_h$ the hexagon pattern is unstable to squares which in turn
for $\chi>\chi_{S}$ are not saturated by the fourth order term in the energy.
In that case within our
perturbation ansatz we are not able to predict which pattern will show
up. For $\chi>\chi_{O}\approx 0.56$ this transition occurs right at
onset and our analysis of hexagons is therefore restricted to $\chi<0.56$. 
This stability criterion for hexagons was to our knowledge not discussed in
the literature before. It was not found by Silber and Knobloch \cite{Silber},
since with their approach the relative stability between hexagons and 
squares could not be determined. It was also missed by Gailitis \cite{Gailitis}
and by Kuznetsov and Spektor \cite{Spektor} since they considered only the
case $\chi \ll 1$.
  
\section{Ferrofluid layer with arbitrary thickness} 
\label{sec:finite}

The aim of this section is to investigate how the finite depth $d$ of a
ferrofluid layer affects the pattern selection. Only ferrofluids with
relatively small susceptibilities  $\chi<\chi_{S}$ will be considered in order
not to leave the region of applicability of our perturbation approach (see the
discussion above). 

\begin{figure}
\begin{center}
\mbox{\epsfxsize=90mm%
\epsffile[32 36 529 424]{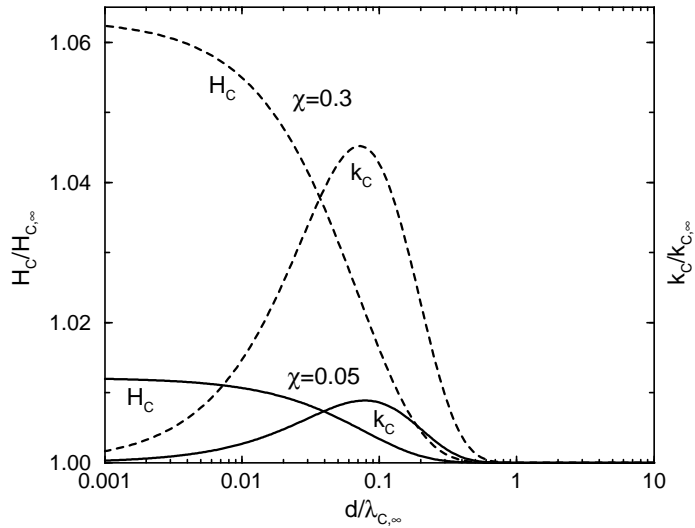}}
\end{center}
\caption{Critical magnetic field $H_{\mathrm{c}}$ and corresponding
critical wavenumber $k_{\mathrm{c}}$ as functions of the layer thickness $d$
in units of $\lambda_{\mathrm{c},\infty}$, where $\lambda_{\mathrm{c},\infty}$
is the critical wavelength for the layer with infinite thickness. Both critical
values are altered less then 2\% for a fluid with $\chi=0.05$ (solid lines)
and less then 7\% for a fluid with $\chi=0.30$ (dashed lines).} 
\label{fig:Hckc}
\end{figure}

 From the linear
stability analysis \cite{Cowley,Weilepp} it is well known that a thin layer
retards the onset of the instability and shifts the  wavenumber of the
unstable mode to larger values. The dependence of the critical magnetic field 
$H_{\mathrm{c},d}$ and of the corresponding critical wavenumber
$k_{\mathrm{c},d}$ on the layer thickness $d$ can be determined from
Eqs.\ (\ref{eq:Hcdkcd}). It is shown in Fig.\ \ref{fig:Hckc}. If the depth $d$
of the fluid is larger then the critical wavelength at infinite depth
$\lambda_{\mathrm{c},\infty}=2\pi/k_{\mathrm{c},\infty}$, which is typically
of the order 
of $10$ mm, the finite thickness of the layer can be ignored. Even
if the layer is very thin, the parameters characterizing the linear instability
of the flat surface are modified only within a few percent for fluids with
small susceptibilities.

\begin{figure}
\begin{center}
\mbox{\epsfxsize=90mm%
\epsffile[62 36 546 424]{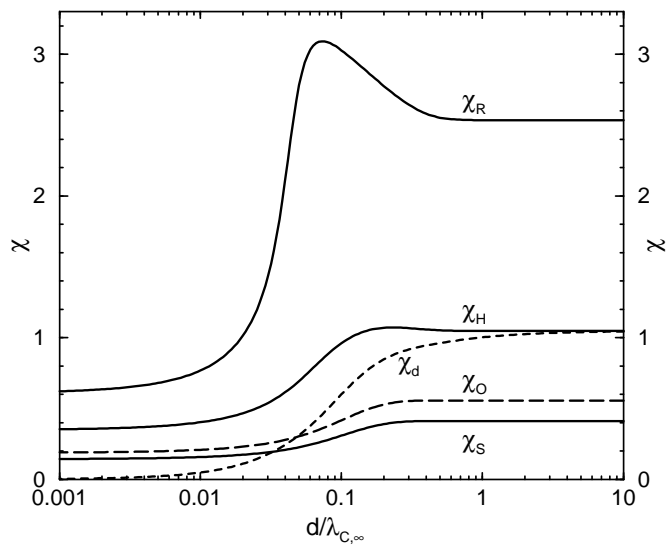}}
\end{center}
\caption{The critical susceptibilities $\chi_R$, $\chi_S$, and $\chi_H$ beyond
  which higher order terms are necessary in the energy functional in order to
  describe the respective pattern consistently as function of the layer
  thickness $d$. Also included are the values for $\chi_{O}$ beyond
  which hexagons are unstable to subcritical squares introduced in section
  \ref{sec:infinite} and $\chi>\chi_{d}(d)$ for which the pattern amplitude
  at onset already exceeds the layer thickness. In both cases our theoretical
  model ceases to be applicable.} 
\label{fig:critical}
\end{figure}

To start with the non-linear analysis we have first to determine the
dependence on $d$ of the critical values $\chi_{R},\chi_{S}$ and $\chi_{H}$
beyond which higher order terms in the energy functional are necessary. The
corresponding results are shown in Fig.\ \ref{fig:critical}. All values of the
critical susceptibilities get eventually smaller if the layer depth
decreases. They do not cross each other such that $\chi_S$ is the smallest one
for all values of $d$. We have included the result for ridges also although
these are unstable in the setup of Fig.\ \ref{fig:setup} for all $d$ and
$\chi$. However, ridges can be stabilized by an additional magnetic field
tangential to the undisturbed surface \cite{Bajaj} and therefore our results
give first informations on the behavior in this modified experimental
setup. 

For very shallow layers the amplitude of the arising pattern may become larger
than the layer depth already at onset. As a consequence the 
ferrofluid layer disintegrates into unconnected regions and ``dry spots''
occur. This topological change is outside our theoretical analysis. The
corresponding value of $\chi$ denoted by $\chi_{d}$ is also shown in
Fig.\ \ref{fig:critical}. As expected it becomes the smallest of all critical
$\chi$-values for $d\to 0$.

In order to investigate the stability of the patterns arising on a layer
with finite thickness we can perform basically the same analysis as in the
previous section. We just have to fix the wavenumber $k$ to its critical value
$k_{\mathrm{c},d}$ corresponding to the given thickness $d$ and to use
the appropriate value for $H_{\mathrm{c},d}$ in the function
$l(\epsilon,k=k_{\mathrm{c},d})$ (see Eq.\ (\ref{eq:l(epsilon,k)})).
Moreover we have to take into account the dependence of the coefficients
$\gamma,g,g_{h},g_{t}$ and $g_{n}$ on the depth $d$. In this way we can
finally determine for any given depth $d$ the stability regions of the
different patterns by means of the corresponding energy function
$f(\{A_n\},k=k_{\mathrm{c},d})$ as given by Eq.\ (\ref{eq:u}). 

\vspace*{0.5cm}
\begin{figure}
\begin{center}
\mbox{\epsfxsize=100mm%
\epsffile[13 13 574 476]{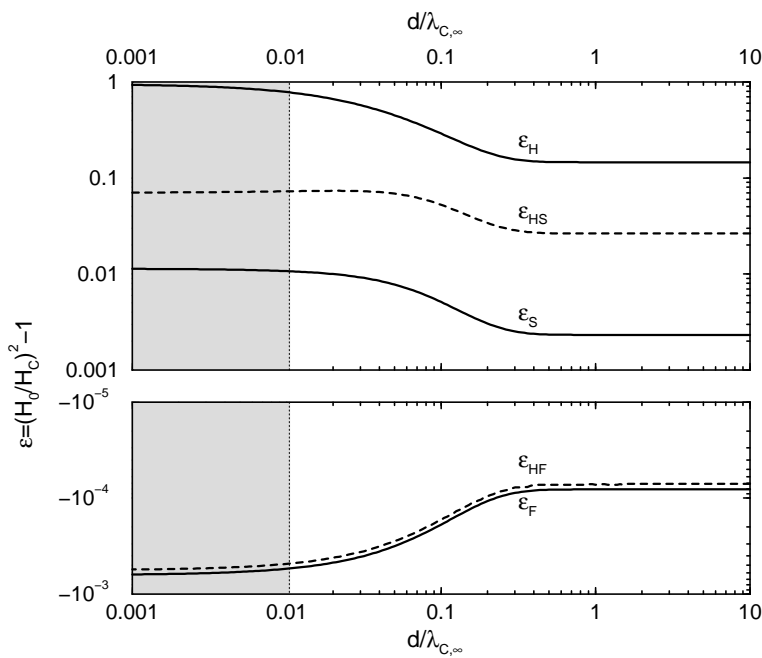}}
\end{center}
\caption{Limits of the stability regions (solid lines) and Maxwellpoints
(dashed lines) as functions of the layer thickness $d$ for a ferrofluid with
$\chi=0.05$. Ridges are always unstable, squares are stable for
$\epsilon>\epsilon_{S}$ and hexagons are stable for 
$\epsilon_{F}<\epsilon<\epsilon_{H}$. In the
gray region the instability results in the formation of dry spots right at
onset, so that our theory no longer applies. The lower part of the figure
describes the subcritical region.} 
\label{fig:chi0.05}
\end{figure}

The results for the various stability boundaries and corresponding Maxwell
points for a fluid with susceptibility $\chi=0.05$ are shown in
Fig.\ \ref{fig:chi0.05}. All the critical values of $\epsilon$ at which
transitions between different planforms arise increase in absolute value with
decreasing layer thickness $d$.
Simultaneously the hysteretic behaviour becomes more pronounced.
As in the linear analysis we find that the effects of 
finite depth can be neglected, as long as the fluid layer is thicker then 
the critical wavelength $\lambda_{\mathrm{c},\infty}$. However, for a thin
layer the stability of the hexagons and squares is changed noticeably.
For instance the value of $\epsilon_{H}$ at which the transition
from hexagons to squares takes place may increase by a factor of $3$ implying
an increase of the magnetic field of about $10\%$. As discussed above for very
thin layers our theory is no longer applicable since the dips of the hexagonal
pattern at onset already reach the bottom of the fluid. The corresponding
values of $d$ are indicated by the shading in Fig.\ \ref{fig:chi0.05}.

\vspace*{0.5cm}
\begin{figure}
\begin{center}
\mbox{\epsfxsize=100mm%
\epsffile[13 13 574 476]{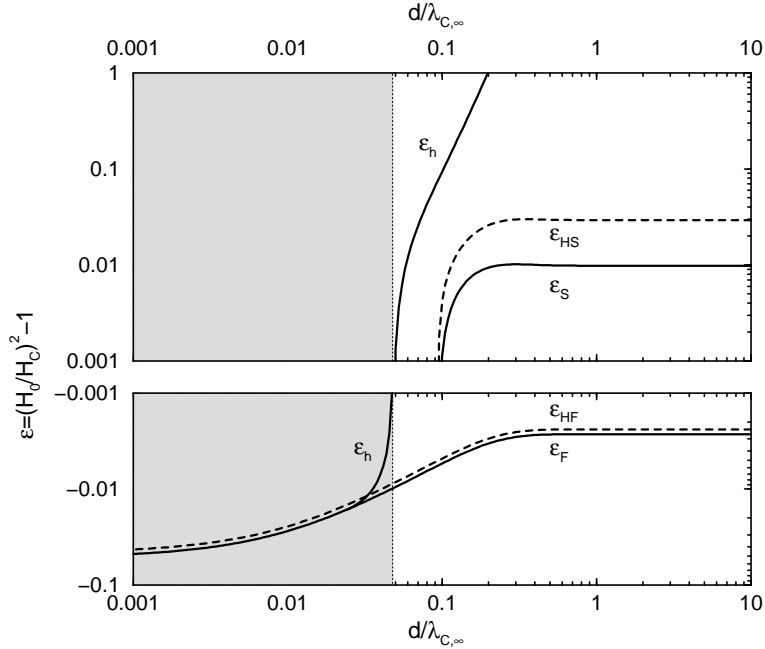}}
\end{center}
\caption{Same as Fig.\ \ref{fig:chi0.05} for a ferrofluid with $\chi=0.30$. With
  decreasing depth $d$ of the layer both $\chi_S(d)$ and
  $\chi_{O}(d)$ become smaller than the susceptibility of the fluid
  higher order terms in the energy would be needed to determine the pattern
  stability.}
\label{fig:chi0.30}
\end{figure}

Fig.\ \ref{fig:chi0.30} displays the analogous results as obtained for a
ferrofluid with the larger susceptibility of $\chi=0.30$. Whereas the
situation for $\epsilon<0$ is rather similar to the one shown in
Fig.\ \ref{fig:chi0.05}, the behaviour for positive $\epsilon$ is qualitatively
different. This is due to 
the fact that although $\chi_S(d\to\infty)>\chi$ with decreasing depth $d$
both $\chi_S(d)$ and $\chi_{O}(d)$ eventually become smaller than 
$0.30$ (cf. Fig.\ \ref{fig:critical}). This gives rise to the divergencies in the
energy discussed above and changes the stability chart accordingly.

\section{Wavenumber selection}
\label{sec:wavenumber}

In many pattern forming systems the wavelength of the first unstable mode is
often a good estimate for the typical length scale of the emerging pattern. If
a whole band of modes is unstable the one with maximal growth rate is likely
to dominate the arising structure \cite{Lange}. On the other hand the
developed pattern is largely determined by the non-linearity in the problem and
therefore its length scale may also substantially deviate from the relevant
scales of the linearized theory. 

A particularly gratifying feature of our variational approach to pattern
formation in the normal field instability in ferrofluids is the possibility to
include the wave number modulus $k$ into the set of parameters varied to
minimize the energy functional. This allows to directly determine the
dependence of the pattern periodicity on the external magnetic field and to
investigate the influence of the variation in wavenumber on the stability
of the different patterns. 

We therefore investigate in the present section the wavenumber selection
problem for the Rosensweig instability. For simplicity we will consider a
ferrofluid with infinite depth only. As shown in the previous section this is
a very good approximation as long as $d>\lambda_{\mathrm{c},\infty}$. We will
compare our findings with recent experiments in which this condition is met. 

The necessary calculations are similar to those reported in section
\ref{sec:infinite} with the only extension that we no longer fix the
wavenumber modulus $k$ in advance but include it into the set of parameters
with respect to which the energy is to be minimized.

\begin{figure}
\begin{center}
\mbox{\epsfxsize=100mm%
\epsffile[3 17 552 421]{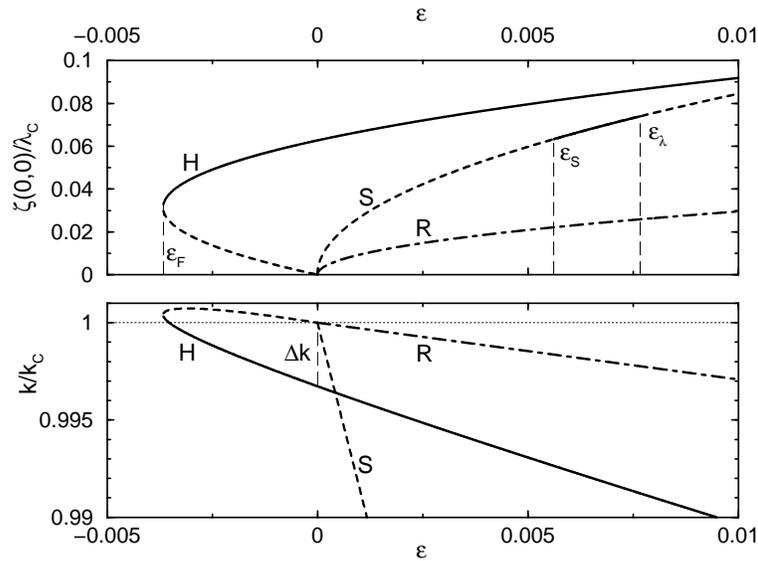}}
\end{center}
\caption{Surface deflection $\zeta(0,0)$ at the cusps of the pattern and
emerging wavenumber $k$ in dependence on the supercriticality
parameter $\epsilon$. For a fluid with $\chi=0.35$ squares are only stable
in the region $\epsilon_{S}<\epsilon<\epsilon_{\lambda}$. Ridges are always
unstable.} 
\label{fig:zetak}
\end{figure}

A typical result is shown in Fig.\ \ref{fig:zetak} displaying the maximal 
surface deflection $\zeta(0,0)$ and the corresponding wavenumber $k$ 
of the patterns. As can be seen from the figure, the wavenumber of the
patterns decreases with increasing supercriticality. This is in
particular pronounced for the square pattern and implies that in this case the
system may lower its energy very efficiently by increasing the wavelength
of the pattern. 

\begin{figure}
\begin{center}
\mbox{\epsfxsize=90mm%
\epsffile[20 36 507 455]{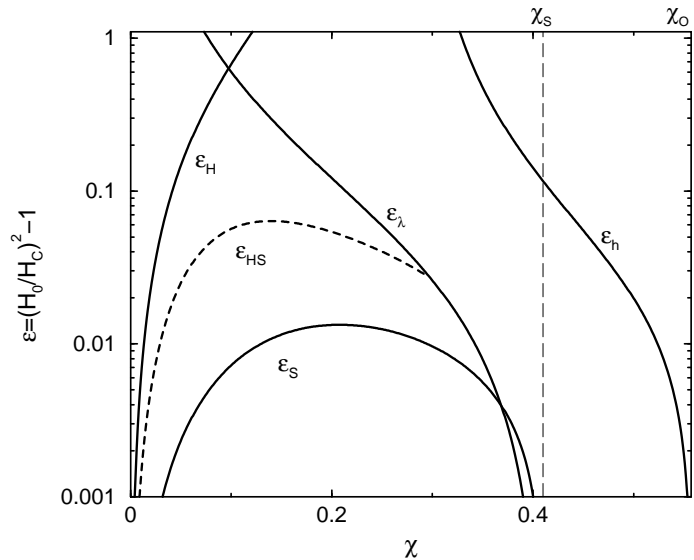}}
\end{center}
\caption{Limits of the stability regions (solid lines) and Maxwell points
(dashed lines) in the $\chi-\epsilon$ plane if variations of the wavenumber
modulus $k$ are taken into account. Squares are stable if
$\epsilon_{S}<\epsilon<\epsilon_{\lambda}$. Hexagons are stable if 
$\epsilon_{F}<\epsilon<\min(\epsilon_{H},\epsilon_{h})$. For large values of
$\chi$ the stability region for squares gets smaller due to the wavelength
instability occurring for $\epsilon>\epsilon_{\lambda}$.}
\label{fig:stabsquares}
\end{figure}

This effect introduces an additional threshold for $\epsilon$ into the
stability chart shown in in Fig.\ \ref{fig:stabsquares}. If 
$\epsilon>\epsilon_{\lambda}$ we find $k\to 0$ and the pattern disappears
by increasing its wavelength to infinity. This process is accompanied by an
unbounded increase of the amplitudes $A_{1},A_{4}$ which carries us outside of
the validity of our perturbation ansatz (\ref{eq:pertan}). The stability
regions of the other pattern remain qualitatively the same as without
minimization in $k$ as can be seen from a comparison between figures 
\ref{fig:stabsquares} and \ref{fig:stability}. 

\begin{figure}
\begin{center}
\mbox{\epsfxsize=90mm%
\epsffile[44 36 507 433]{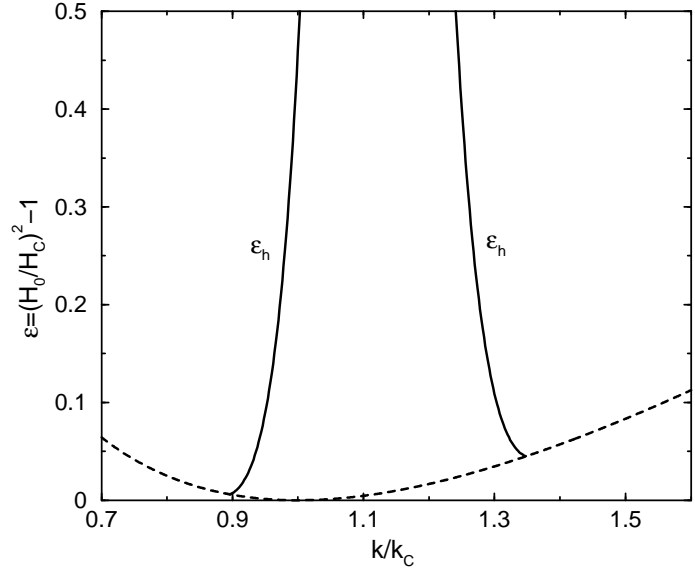}}
\end{center}
\caption{Stability band of the hexagonal pattern in the $k-\epsilon$ plane
for a ferrofluid with $\chi=0.35$. Hexagons become unstable when 
$\epsilon>\epsilon_{h}$. 
The marginal stability curve (dashed line) is also displayed.}
\label{fig:band}
\end{figure}

For supercritical magnetic fields a whole band of wavenumbers giving rise to
stable patterns exists. For hexagons this stability region is shown in Fig.\
\ref{fig:band}. 
At fixed $\epsilon$ the hexagonal planform  becomes unstable if 
the wavenumber is increased sufficiently. In \cite{Abou} such an increase of
the wavenumber was achieved experimentally by compressing the hexagonal 
pattern in a hopper. The observed transition from a hexagonal to a square
pattern with smaller wavenumber occurs when the limit $\epsilon_{h}$ of the
stable $k$-band is exceeded and is hence in qualitative agreement with our
theory. 

\section{Discussion}
\label{sec:discussion}

In the present paper we have theoretically investigated the formation of
patterns on the free surface of a ferrofluid resulting from the Rosensweig
instability. For a ferrofluid layer of arbitrary depth $d$ in an external
magnetic field ${\bf H}_{0}$ the equilibrium surface profile was determined by
minimizing the appropriate thermodynamic potential. We have assumed a linear
dependence of the magnetization on the magnetic field which is at most 5\% off
the correct values for the experimentally situations with which we compare our
findings. The analysis is
restricted to the vicinity of the critical magnetic field at which the flat
interface becomes unstable since it uses a perturbation expansion for  the
energy up to fourth order in the amplitude of the surface deflection. 
Our findings generalize the classical results of Gailitis \cite{Gailitis} and
Kuznetsov and Spektor \cite{Spektor} to layers of finite depth
$d<\infty$. 

We have found that, at least for the slightly supercritical magnetic fields
considered, the finite thickness of the ferrofluid layer can be neglected
as long as it remains larger than the critical wavelength of the linear 
instability which in typical experiments is about 1 cm. For thinner layers
nonlinear effects become increasingly important. This shows up quite generally
in more pronounced hysteresis effects for the transition between different
planforms. We also find smaller values of the susceptibilities for which
squares and ridges may appear through backward bifurcations and correspondingly
higher order terms in the energy are necessary to saturate the instabilities.
The critical values of the susceptibility $\chi$ beyond which our theory breaks
down are displayed in Fig.\ \ref{fig:critical}. Together with the findings
shown in Fig.\ \ref{fig:validation} they quantify the qualitative
statement $\chi\ll 1$ used in \cite{Gailitis,Spektor}. An extension of our
analysis including higher orders similar to what has been done in \cite{Engel}
for the one-dimensional situation seems feasible but rather tedious. 

It is possible to understand the enhanced hysteretic behaviour for thin layers
qualitatively. A thin layer inhibits surface deformations resulting
from the Rosensweig instability (see Fig.\ \ref{fig:Hckc}) since the magnetic
field energy is suppressed due to the finite depth of the layer. On the other
hand for the developed pattern the magnetic flux is concentrated at the peaks
of the surface so that the effective layer depth becomes larger.
Correspondingly the
pattern is additionally stabilized and decays only after a reduction of the
field stronger than expected from the linear theory. A similar reasoning
explains why backward bifurcations of ridges and squares are facilitated in
thin layers. 

In addition to determining the stable planforms by minimizing the energy
functional in the amplitudes of the different modes present in our ansatz for
the surface profile, we have also addressed the problem of wavenumber selection
by including the wavenumber modulus $k$ into the set of variational
parameters. We found quite generally a decrease of the wavenumber of the
pattern with increasing magnetic field. This is in qualitative agreement with
recent experiments by Abou {\it et al.} \cite{Abou} in which the the field
intensity was abruptly increased to overcritical values and the corresponding 
wavelength $\lambda=2\pi/k$ of the resulting hexagonal pattern was found to
increase with increasing field. Likewise it was found that the wavenumber $k$
of the hexagonal pattern is already for slightly supercritical magnetic
fields smaller then the critical wavenumber $k_{\mathrm{c}}$. The
corresponding difference $\Delta k$ is also found in the theory and shown in 
Fig.\ \ref{fig:zetak}. Unfortunately a quantitative comparison with the
experimental results is impossible since in the experiment a ferrofluid with a
susceptibility of $1.4>\chi_{H}$ was used. For such a susceptibility the
theoretical analysis would require higher order terms in the perturbation
expansion of the energy. Nevertheless
we found that $\Delta k$ grows with increasing susceptibility such that the
experimentally observed value of a few percent agrees well with our result
for $\Delta k$ at smaller susceptibilities $\chi<\chi_{H}$. 

Moreover the theoretical analysis has shown that when including the wavenumber
into the set of variational parameters the square pattern gets rather
susceptible to an unbounded increase in the wavenumber. This gives rise to a 
reduced stability domain for squares and may be related to the fact that in
the experiments using jumps in the field intensity always hexagonal arrays of
peaks were found \cite{Abou}. Quite generally our results verify that the
resulting planform may depend on the details of the experimentally chosen way
to reach overcritical magnetic fields.

Let us finally stress that the variation of the wavenumber of the patterns
with the external field as found here theoretically and observed in the
experiments does not invalidate our ansatz with a fixed wavenumber used
before. As reported in \cite{Abou} the wavenumber of the arising hexagonal
pattern coincides with $k_{\mathrm{c}}$ if the magnetic field is increased
and decreased in a quasi-static way. This is probably due to the fact that the
boundary conditions suppress the emergence or disappearance of peaks in a
developed pattern. Therefore using the quasi-static process a metastable
pattern with $k_{\mathrm{c}}$ is produced whereas a jump in field intensity
gives rise to the most stable pattern corresponding to a smaller wavenumber.

\acknowledgments
We would like to thank B\'ereng\`ere Abou for explaining her experimental
findings to us and Adrian Lange for interesting discussions. This
work was supported by the {\it Deutsche Forschungsgemeinschaft} under the
project En 278/2-1.


\end{document}